\documentclass[aps,pre,reprint,superscriptaddress]{revtex4-2}

\usepackage{bm,amsmath,mathtools,amsfonts}
\usepackage{siunitx}
\usepackage{graphicx, color}

\usepackage{tikz}
\usetikzlibrary{arrows.meta, patterns, decorations.pathmorphing, decorations.markings, calc}

\def\arr{{>[length=0.1cm]}}
\definecolor{myorange}{RGB}{253,141,60}
\definecolor{myblue}{RGB}{62, 95, 239}
\definecolor{myred}{RGB}{214,39,40}

\begin{document}


\title{Dynamic clustering of active chemotactic Janus particles}


\author{Joël Mabillard}
\affiliation{Université Paris Cité, CNRS, Matière et Systèmes Complexes, F-75013 Paris, France}
\author{Frédéric van Wijland}
\affiliation{Université Paris Cité, CNRS, Matière et Systèmes Complexes, F-75013 Paris, France}
\affiliation{Yukawa Institute for Theoretical Physics, Kyoto University, Kyoto, 606-8502, Japan}
\author{Jean-Baptiste Fournier}
\affiliation{Université Paris Cité, CNRS, Matière et Systèmes Complexes, F-75013 Paris, France}


\date{\today}

\begin{abstract}
Modeling self-propulsion as an emergent diffusiophoretic effect, we come back to the dynamic clustering experiment of active Janus colloids. Working in the original three-dimensional geometry, and using the chemical solute concentration as a control parameter, we find that the dynamic clustering region occupies a robust fraction of the phase diagram.  This behavior is rationalized using a linear stability analysis that efficiently captures the instability of the gas phase to dynamic clusters. We also find that dynamic clustering results from the combined effects of a Keller-Segel collapse and of a motility-induced phase separation, either being arrested by the orientational coupling to the chemical field in agreement with previous work.
\end{abstract}

\maketitle

\section{Active matter, Janus colloids and dynamical clustering}
\label{sec:intro}

Active particles, such as energy-transducing proteins, self-propelled colloids, bacteria, and animal collectives, operate far from equilibrium---even in a steady state---due to continuous energy consumption at the particle level~\cite{marchetti_hydrodynamics_2013,gompper_2020}. They exhibit fascinating behaviors that have no counterpart in passive Brownian systems. For instance, active motile particles tend to accumulate in dense regions where they move more slowly, leading to motility-induced phase separation (MIPS) in the absence of attractive forces~\cite{cates_motility-induced_2015}.

In the realm of colloidal physics, active particles can be physically realized through diffusiophoresis---a process where fluid flow is induced past a surface by solute concentration gradients~\cite{anderson_colloid_1989,golestanian_phoretic_2018}. Indeed, as demonstrated by Golestanian {\it et al.}, colloids can propel via self-diffusiophoresis when a chemical reaction is catalyzed asymmetrically over a colloid’s surface~\cite{golestanian_propulsion_2005, golestanian_designing_2007}. This phenomenon is experimentally realized in Janus colloids \cite{theurkauff_dynamic_2012,palacci2013living,singh2017non,ginot_aggregation-fragmentation_2018}. Such active particles undergoing phoretic interactions can form dynamic clusters~\cite{theurkauff_dynamic_2012,saha_clusters_2014,palacci2013living,singh2017non,ginot_aggregation-fragmentation_2018,pohl_dynamic_2014,liebchen_phoretic_2017,liebchen2019interactions,panja2026boundary} characterized by continuous rearrangements or undergo collective collapse, analogous to microorganisms interacting via chemical signaling through the Keller-Segel (KS) instability~\cite{keller_initiation_1970,keller_model_1971}.

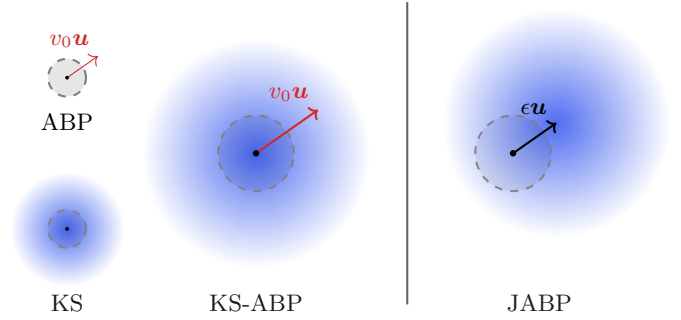
\begin{figure}[t!]
    \centering
\begin{tikzpicture}
\begin{scope}[shift={(0, 2)}]
        \filldraw[draw=gray, thick, dashed, fill=gray!20] (0,0) node[below=0.35cm] {ABP} circle (.25);
        \draw[-\arr, myred] (0,0) -- (35:0.5) node[above left] {$v_0\bm{u}$};
        \filldraw (0,0) circle (0.5pt);
    \end{scope}
   
\begin{scope}[shift={(0, 0)}]
        \fill[fill=gray] (0,0) circle (.25);
        \path[shading=radial, inner color=myblue, outer color=white, opacity=.9] (0,0) circle (0.75);
        \filldraw (0,0) circle (0.5pt);
        \draw[draw=gray, thick, dashed, opacity=0.9] (0,0) node[below=.75cm] {KS} circle (.25);
    \end{scope}

\begin{scope}[shift={(2.5, 1)}]
        \fill[fill=gray] (0,0) circle (.5);
        \path[shading=radial, inner color=myblue, outer color=white, opacity=.9] (0,0) circle (1.5);
        \draw[draw=gray, thick, dashed, opacity=0.9] (0,0) node[below=1.75cm] {KS-ABP} circle (.5);
        \draw[-\arr, thick, myred] (0,0) -- (35:1) node[above left] {$v_0\bm{u}$};
        \filldraw (0,0) circle (1pt);
    \end{scope}
    
\draw[draw=black!60, thick] (4.5, -1) -- (4.5, 3.);    
    
\begin{scope}[shift={(5.9, 1)}]
        \fill[fill=gray] (0,0) circle (.5);
        \path[shading=radial, inner color=myblue, outer color=white, opacity=.9] (35:0.7) circle (1.5);
        \filldraw (0,0) circle (1pt);
        \draw[draw=gray, thick, dashed, opacity=0.9] (0,0) node[below=1.75cm, xshift=0.35cm] {JABP} circle (.5);
        \draw[-\arr, thick] (0,0) -- (35:0.7) node[above left] {$\epsilon\bm{u}$};
    \end{scope}

\end{tikzpicture}
\caption{Schematic representation of  four particle models: Active Brownian Particle (ABP, upper left), Keller-Segel particle (KS, lower left), Keller-Segel Active Brownian Particle (KS-ABP, center), and our Janus Active Brownian Particle (JABP, right). The dashed circle delineates the excluded volume boundary of the particle core. Here, $v_0$ denotes the self-propulsion velocity along the orientation vector $\bm{u}$, and $\epsilon$ represents the spatial eccentricity of the chemical source. The shaded color gradient illustrates the profile of the chemical concentration field $c$.}
    \label{fig:models}
\end{figure}

The modeling of assemblies of self-diffusiophoretic particles is usually based on the combination of two {\it a priori} independent ingredients, as shown in Fig.~\ref{fig:models}.  The first one is self-propulsion, which is implemented using constant-velocity particles whose orientation undergoes rotational diffusion, known as active Brownian particles (ABP)~\cite{romanczuk_active_2012}. This model emulates the local catalysis and diffusiophoretic propulsion. The second ingredient is a chemotactic interaction mediated by a concentration field $c$. This can emanate from the center of the particle~\cite{pohl_dynamic_2014,panja2026boundary}, in a Keller-Segel fashion, or, as an improvement, from a hemisphere around the particle~\cite{saha_clusters_2014, liebchen_phoretic_2017,liebchen2019interactions}. In these and in subsequent works, the authors find a wealth of dynamic phases. The dynamic cluster phase found in \cite{pohl_dynamic_2014} is of particular relevance to the description of the experiments~\cite{theurkauff_dynamic_2012}.

In this work we adopt the realistic viewpoint that in Janus particles the same physics and chemistry govern both self-propulsion and chemotactic interactions. Of course, this means extending the validity of the far field expressions for the velocity field down to the particle scale, as was also done in~\cite{pohl_dynamic_2014, liebchen_phoretic_2017,liebchen2019interactions,panja2026boundary} for chemotactic interactions within dense clusters. In our approach, this is the sole ingredient driving the self-propulsion. We shall  establish that this leads to the exact same expression for the self-propulsion velocity as that found in \cite{golestanian_designing_2007}. In our viewpoint, the interplay between self-propulsion and chemotactic interactions is instrumental in faithfully describing  the formation of dense clusters, where the chemical field is depleted. This fundamental ingredient is carefully described and critically discussed in Section~\ref{sec:JABPmodel} and cast in a continuum description in terms of the chemical and particle concentration fields. In order to better approach the experimental setting of \cite{theurkauff_dynamic_2012} we use, in Section~\ref{sec:Model_exp}, the geometry in which the active colloids live on a two-dimensional substrate while embedded in a three-dimensional solute. Conducting a linear stability analysis shows the existence of a transition to dynamic clustering. Performing extensive numerical simulations shows that there always exists a dynamic cluster regime in between the gas and collapsed state. A few words  on the respective roles of the KS instability and MIPS in achieving dynamic clustering are gathered in the conclusion in Section~\ref{sec:conclustion}. 

\section{The Janus-ABP model}
\label{sec:JABPmodel}


We now introduce the active particle model that exhibits the specific interplay between self-propulsion and chemotaxis discussed in our introduction. We refer to this model as 
a Janus active Brownian particle (JABP). We first describe a single particle and subsequently a many-particle system.

\subsection{One particle}

We consider a particle in dimension $d$ with position $\bm r(t)$ and orientational unit vector $\bm u(t)$, interacting with a solute field $c(\bm x,t)$ according to
\begin{align}
\partial_t\bm r&=\mu\bm\nabla c(\bm r)+\sqrt{2 D_t}\,\bm \xi(t),\label{eq:motion-trans}\\
\partial_t\bm u&=\sqrt{2 D_r}\bm\eta^\perp(t),\\
\partial_tc&=D\bm\nabla^2c + k\delta^{(d)}(\bm x-\bm r-\varepsilon\bm u),
\end{align}
where $\bm\eta^\perp(t)=(\bm I-\bm u\otimes\bm u)\bm \eta(t)$ produces a random rotation of the particle's orientation. Both $\bm \eta$ and $\bm \xi$ are Gaussian white noises with independent components. Parameters-wise, $D_t$ and $D_r$ are the translational and rotational diffusion constants of the colloid, respectively, and $\mu$ is  its  phoretic  mobility. The solute diffuses with a constant $D$ and $k$ is its production or consumption rate (depending on its sign).

The new features in these equations with respect to Refs.~\cite{pohl_dynamic_2014, liebchen_phoretic_2017,liebchen2019interactions,panja2026boundary} are not only the absence of an autonomous self propulsion term $v_0\bm u$ in Eq.~\eqref{eq:motion-trans}, but also the presence of the $\varepsilon>0$ parameter. The solute field produced by the source term centered at $\bm r+\varepsilon\bm u$, on the side of the catalytic hemisphere, when combined with $\mu\bm \nabla c$,  generates the particle's velocity. Thus, like a Janus colloid, this particle is set into motion by the gradient of solute it creates a small distance ahead of its center (see the right panel of Fig.~\ref{fig:models}). 

In principle, the term $\mu\bm\nabla c$ accurately represents the phoretic velocity of a colloid only when the concentration gradient is uniform at the scale of the particle~\cite{anderson_colloid_1989}. Here, however, the concentration is highly heterogeneous due to the vicinity of the chemical sink produced by the colloid itself. We shall nevertheless resort to that description, since, as we show next, we obtain a good match with the exact autophoretic velocity obtained in~\cite{golestanian_designing_2007} from a detailed microscopic calculation.

Consequently, in the presence of many particles, using the term $\mu\bm\nabla c$ is a physically relevant approximation not only for self-propulsion and far-field phoretic interactions, but also, qualitatively, for near-field interactions within particle clusters. 

\subsubsection{The ABP limit}

Let us consider the fast solute limit, {\it i.e.} $D\gg D_t$ (which we will use throughout).  In this regime,  $c$ instantaneously equilibrates thus we set $\partial_t c=0$ and we obtain  $c(\bm x)=-(k/D)G_d(\bm x-\bm r-\varepsilon\bm u)$, where $G_d$ is the Laplacian's Green function in $d$ dimensions. Then $\mu\nabla c(\bm r)=(k\mu/D)G'_d(\varepsilon)\bm u$ and the particle is effectively an ABP, evolving according to
\begin{align}
\partial_t\bm r&=v_0\bm u+\sqrt{2 D_t}\,\bm \xi(t),\\
\partial_t\bm u&=\sqrt{2 D_r}\bm\eta^\perp(t),
\end{align}
with
\begin{equation}
v_0=\frac{k\mu}{D}G'_d(\varepsilon)
=
\begin{cases}
\dfrac{k\mu}{2\pi D\,\varepsilon}, & (d=2),\\[2ex]
\dfrac{k\mu}{4\pi D\,\varepsilon^2}, & (d=3).
\end{cases}\label{eq:v0}
\end{equation}
When the production/consumption rate $k$ depends on $c$, we thus expect that $v_0$ also depends on $c$ (and thus, indirectly, in a many body problem, on the local colloidal density). This velocity compares well with the exact velocity $v_0=\mu\alpha/(4D)$ obtained in \cite{golestanian_designing_2007}  (their Eq.~(9)) for a hemispherical Janus colloid of radius $a$ in three dimensions. In their derivation, $\mu$ is the average phoretic mobility along the surface of the particle, and $\alpha=k/(2\pi a^2)$ the surface activity of the catalytic half. The match is exact for the effective choice  $\varepsilon=a\sqrt{2}$. Thus, for our Janus-ABP particle, $\varepsilon$ plays the role of the particle's radius. As discussed above, this result justifies using the $\mu\bm\nabla c$ term for both self-propulsion and phoretic interactions in many-particle systems. 

\subsection{Many particles}

Next, we consider $N$ particles. By analogy to Janus colloids, we introduce a torque with coefficient $\mu'$ that aligns $\bm u$ with $\bm\nabla c$ (or oppose it, depending on the sign of $\mu'$). The dynamics of the particles and the field is thus given by
\begin{align}
\partial_t\bm r_n(t)&=\mu\bm\nabla c(\bm r_n)+\sqrt{2 D_t}\,\bm \xi_n(t),\label{eq:rn2d}\\
\partial_t\bm u_n(t)&=
(\bm I-\bm u_n\otimes\bm u_n)\left(\mu'\bm\nabla c(\bm r_n)+
\sqrt{2 D_r}\bm\eta_n\right),\label{eq:thetan2d}\\
\partial_tc(\bm r,t)&=D\bm\nabla^2c + k\sum_{n=1}^N\delta^{(d)}(\bm r-\bm r_n-\varepsilon\bm u_n)\label{eq:c2d}.
\end{align}
Because $\bm\nabla c(\bm r_n)$ is generated by all particles, the system exhibits both self-propulsion and chemotactic interactions, similarly to the earlier models in \cite{pohl_dynamic_2014, liebchen_phoretic_2017}. The main difference is that in our case the particles do not have a constant propulsion speed; instead, their speed is an emergent property influenced by the presence of neighboring particles, much like in real colloids.

Let us put this model into perspective with what is known from more microscopic descriptions~\cite{anderson_colloid_1989}. In the presence of a uniform solute gradient, $(\bm\nabla c)_\infty$, the translational velocity $\bm V$ and the angular velocity $\bm\Omega$ are given by $\bm V=\bm\mu(\bm\nabla c)_\infty$ and $\bm\Omega=\bm\mu'\times(\bm\nabla c)_\infty$ (in three dimensions). The tensor $\bm\mu$ and the vector $\bm\mu'$ are given by~\cite{anderson_colloid_1989}
\begin{align}\label{eq:Anderson}
    \bm\mu=
\langle \zeta\rangle\bm I
-\frac32\left\langle\left(\bm n\otimes\bm n-\frac13\bm I\right)\zeta\right\rangle,
\quad
\bm\mu'=\frac9{4R}\langle{\zeta\bm n}\rangle,
\end{align}
where $\zeta(\bm r)$ is the local surface phoretic coefficient (the linear coefficient relating the local concentration gradient and the local slip velocity of the fluid past the surface), $\langle\ldots\rangle$ is the surface average and $\bm n$ the local vector normal to the surface. It also follows from Eq.~\eqref{eq:Anderson} that $\mu$ and $R\mu'$ must be of the same order of magnitude for a Janus colloid, which we shall assume in the following. 

Note that for a half-coated spherical colloid, the anisotropic contribution to $\bm\mu$ vanishes entirely, yielding $\bm V = \mu(\bm\nabla c)_\infty = \langle\zeta\rangle(\bm\nabla c)_\infty$ regardless of the particle's orientation. Thus, the chemotactic term $\mu\bm\nabla c(\bm r_n)$ is correct for Janus colloids. As we have mentioned before, strictly speaking it is accurate only when the particles are well separated for the solute gradient to be uniform. However, following Ref.~\cite{pohl_dynamic_2014, liebchen_phoretic_2017,liebchen2019interactions,panja2026boundary}, we shall adopt this expression as a simplifying approximation even within dense particle clusters. 

Furthermore, unlike the KS-ABP model (see Fig.~\ref{fig:models}), where the chemical source or sink is located at the particle center, our JABPs consume the chemical at a position displaced by $\epsilon\bm u$, mimicking the asymmetric catalytic surface of a Janus colloid. As shown in~\cite{saha_clusters_2014,liebchen_phoretic_2017}, this finite extension proves important when particles are in close contact, where the chemical field varies over the scale of the particle size.

\section{Modeling real Janus colloids}
\label{sec:Model_exp}
In this section, we model the experiments of Theurkauff {\it et al.}~\cite{theurkauff_dynamic_2012} using the JABP model. They consider a dense active suspension of Janus colloids  sedimented under gravity at the bottom of a three-dimensional cell containing a hydrogen peroxide solution. These active particles propel via self-phoretic motion powered by the hydrolysis of water peroxide. The authors observed the formation of a dynamic ``cluster phase'' where the typical cluster size increases linearly with the Péclet number characterizing the self-propulsion velocity (the size of a cluster is defined as the number of particles it contains). We begin by endowing the JABP particles with additional ingredients to take into account the features of the experiment. Then we rely on a description in terms of particle density and polarization fields to perform a linear stability analysis to detect the onset of the dynamic cluster phase. Finally we present extensive numerical simulations revealing the extent of the dynamic cluster phase arising from the emergent self-propulsion velocity. 

\subsection{Model}
 The chemical solute concentration $c(\bm R, t)$ is defined throughout the bulk, where $\bm R = (\bm r,z)$ is a three-dimensional vector, $\bm r\in[0,L_x[\times[0,L_y[$ and $z\in[0,L_z]$. We model the $N$ particles as being confined at the bottom of the box $(z=0)$, with positions $\bm r_n$ and orientations $\bm u_n$, parameterized by the polar $\theta_n$, lying in a two-dimensional plane. We now adapt the JABP model described in Sec.~\ref{sec:JABPmodel}. The steric interactions between the colloids are taken into account by means of an excluded-volume term $\bm f_{\rm ex}(\{\bm r_k\})$ acting on each colloid. The colloids consume the solute at a local rate governed by Michaelis-Menten kinetics~\cite{howse_self-motile_2007}, according to which $k = -h c$, where $h$ denotes the catalytic activity and $c(\bm{R}, t)$ is the local concentration. The diffusion coefficient $D(\bm R,\{\bm r_k\})$ of the chemical field is space-dependent through the colloids' positions, reflecting the fact that diffusion is hindered by the particles, especially in dense regions. These features specific to the experiment lead to the modified equations
\begin{align}\label{eqa}
\partial_t\bm r_n&=\mu\bm\nabla c(\bm r_n)+ \bm f_{\rm ex}(\{\bm r_k\})+\sqrt{2 D_t}\,\bm \xi_n(t) \,,\\\label{eqb}
\partial_t\theta_n&=\mu'\bm u^\perp_n\cdot\bm\nabla c(\bm r_n)  +\sqrt{2 D_r}\,\eta_n(t)\,,\\
\partial_t c&= \bm\nabla\cdot\left(D(\bm R) {\bm\nabla} c\right)
- hc\sum_n\delta^{(3)}(\bm R-\bm r_n-\epsilon\bm u_n).
\label{eqc}
\end{align}
The chemical concentration is maintained at a fixed value $c_\infty$ at the top boundary $z=L_z$ by a reservoir, imposing a Dirichlet boundary condition. At the bottom of the cell, a vertical no-flux boundary condition $\partial_z c(z=0)=0$ is imposed. 

Due to the Michaelis-Menten mechanism, our emergent self-propulsion velocity is now proportional to the local solute concentration, in agreement with the experiment~\cite{theurkauff_dynamic_2012}, which contrasts with previous models of phoretic Janus colloids~\cite{pohl_dynamic_2014,liebchen_phoretic_2017}. In addition the configuration-dependent diffusion constant of the solute participates in the reduction of the propulsion velocity inside dense clusters, where the local concentration of chemical fuel is depleted.  Therefore, when the fuel is locally exhausted, the particle activity naturally decreases and can eventually stop, as expected in dense aggregates.

\subsection{Description in terms of fields}
The colloids live in the two-dimensional plane at the bottom of the experiment. As shown in Appendix~\ref{app:DK},  the particles' fluctuating density in positional and angular space,
\begin{align}
    \rho(\bm r,\theta,t)=\sum_n\delta^{(2)}(\bm r-\bm r_n)\delta(\theta-\theta_n),
\end{align}
evolves according to the Dean-Kawasaki equation:
\begin{align}
    \partial_t\rho &= - \bm \nabla\cdot
    \left(
    \rho\phi\mu\bm\nabla c
    -D_t\bm\nabla\rho
    -\sqrt{2D_t\phi\rho}\,\bm \xi
    \right)
    \nonumber\\
    &-\partial_\theta\left(
    \rho\mu'\bm u^\perp
    \cdot\bm\nabla c
    -D_r\partial_\theta\rho
    -\sqrt{2D_r\rho}\,\eta
    \right),\label{eq:DK}
\end{align}
where $\bm\xi(\bm r, \theta,t)$ and $\eta(\bm r, \theta,t)$ are centered Gaussian white noises with unit variance. The quantity
\begin{equation}
\bar\rho(\bm r)=\int_0^{2\pi}d\theta\,\rho(\bm r, \theta)
\end{equation}
is the local particle density.
The density of empty space
$\phi=1-\bar{\rho}/{\bar{\rho}_m}$ has been inserted to phenomenologically account for excluded volume effects. It involves the maximum packing density $\bar{\rho}_m$ and it ensures the vanishing the mobility at $\bar{\rho}=\bar{\rho}_m$. By not dressing $\mu'$ by a density-dependent factor, we have assumed that steric interactions do not affect the rotational degrees of freedom, thus expressing that particles can freely reorient even within dense clusters.

We now turn to the equation for the solute concentration. Assuming the relevant physics is caught by the leading order in a vanishing  $\varepsilon$ limit, the equation for $c(\bm R,t)$ in three-dimensional space becomes
\begin{align}
\label{eq:cepsilon}
\partial_t c =  \bm\nabla\cdot(D(\bar{\rho})\bm \nabla c) -hc\delta(z)\bar\rho + hc\delta(z)\varepsilon\bm\nabla\cdot\bm p + O(\varepsilon^2),
\end{align}
where
\begin{equation}
    \bm p(\bm r)=\int_0^{2\pi}d\theta\,\bm u(\theta)\rho(\bm r, \theta),
\end{equation}
is the polarization field. It is worth noting that the same polarization term $\propto\nabla\cdot\bm p$ arises in the solute equations of Refs.~\cite{saha_clusters_2014,liebchen_phoretic_2017}, although their models consider a source distributed over a finite hemisphere. The diffusivity $D(\bar\rho)$ equals $D$ in the bulk ($z>2a$, where $a$ is the radius of the colloid) and $D(1-\bar\rho/\bar\rho_m)$ in the colloidal layer of thickness ($z\le2a)$, thereby accounting for hindered diffusion in the simplest way.

Note also that for $\varepsilon=0$ and in mean-field, the equations describing our JABP particles become the Keller-Segel equations. Indeed, neglecting the noise term in Eq.~\eqref{eq:DK}, and integrating over $\theta$ yields
\begin{align}
      \partial_t\bar\rho &= - \bm \nabla\cdot
    \left(
    \bar\rho\mu\bm\nabla c
    -D_t\bm\nabla\bar\rho\right),
    \nonumber\\
    \partial_t c&=D\nabla^2c + k\bar\rho,
\end{align}
which is the Keller-Segel model. This connection was already noted in Refs.~\cite{theurkauff_dynamic_2012, pohl_dynamic_2014}. Clusters are observed experimentally, and the Keller-Segel limit is know to exhibit some instability, which prompts us to explore the linear stability of the homogeneous phase.

\subsection{Linear stability analysis}

\begin{figure}[t!]\centering
{\includegraphics[width=\columnwidth]{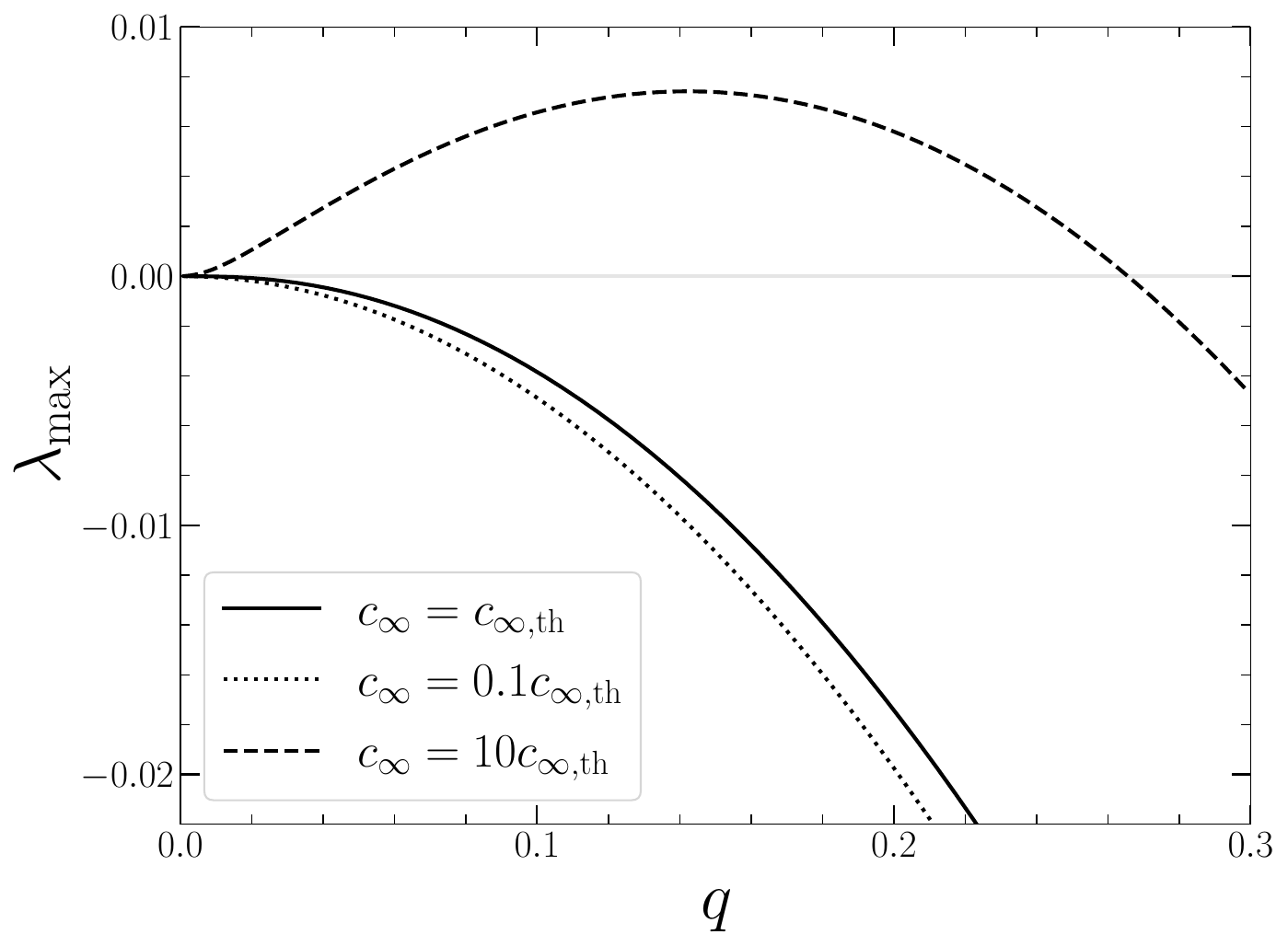}}
\caption[] {Largest eigenvalue $\lambda_{\rm max}(q)$ obtained from the linear stability analysis as a function of the wavenumber $q$. The three curves correspond to $c_\infty=0.1c_{\infty, \rm th}$ (dotted), $c_\infty=c_{\infty, \rm th}$ (solid), and $c_\infty=10c_{\infty, \rm th}$ (dashed), where $c_{\rm th}=13.67\times10^{4}$. The instability threshold is identified by the change of sign of $\lambda_{\rm max}$. All other parameters are identical to those used in the simulations, with $\mu=-10^{-4}$ and $\mu'=10^{-5}$.
} \label{Fig:EV}
\end{figure}

To probe the stability of an isotropic homogeneous state with particle density $\bar\rho_0$, we perform a linear stability analysis of the equations \eqref{eq:DK} and \eqref{eq:cepsilon} with the boundary conditions $c(\bm r,L_z)=c_\infty$ and $\partial_z c(\bm r,0)=0$. The details of the calculation are presented in Appendix~\ref{app:LSA}. {The background concentration profile $c_0(z)$ is found to be linear in both the bulk and the colloidal layer, with a change in slope at $z=2a$ due to the continuity of the diffusive flux at the interface. The bottom value is given by}
\begin{align}
	 c_0(0) & = \frac{c_\infty}{1+h\bar\rho_0\left(\frac{2a}{D\left(1-\bar\rho_0/\bar\rho_m\right)}
	 +\frac{L_z-2a}{D}\right)}.\label{eq:c00}
\end{align}
As expected, if the reservoir is infinitely far away ($L_z$ diverges) or if the bottom layer is filled with colloids ($\bar\rho_0=\bar\rho_m$), the background concentration profile vanishes.

Next, we consider a perturbation in the particle density of the form
\begin{align}
\rho({\bf r},\theta,t) &= \frac{\bar \rho_0}{2\pi}+ \delta \rho({\bf r},\theta,t).
\end{align}
It induces a corresponding variation $\delta c(\bm r, z, t)$ in the solute concentration profile. We expand these perturbations in Fourier modes:
\begin{align}
	\delta \rho({\bf r},\theta,t) &= \int \frac{d^2{\bm q}}{(2\pi)^2}\sum_{m\in\mathbb{Z}}\delta \rho_m({\bm q},t)e^{i({\bm q}\cdot{\bm r}+m\theta)},\\
	\delta c({\bm r},z,t) & = \int \frac{d^2{\bm q}}{(2\pi)^2}\delta c({\bm q},z,t)e^{i{\bm q}\cdot{\bm r}}.\label{eq:FTdc}
\end{align}
 Because solute diffusion is much faster than the characteristic time scale of colloidal motion, we assume that this concentration profile adjusts instantaneously. Thus we set $\partial_t\delta c=0$ in Eq.~\eqref{eq:cepsilon}. Linearizing Eqs.~\eqref{eq:DK}-\eqref{eq:cepsilon} and integrating out the concentration profile, we find that the three modes $\delta\rho_0$, $\delta\rho_1$ and $\delta\rho_{-1}$ are decoupled from the higher-order modes (to the linear order in $\varepsilon$ we work with), and obey
 \begin{align}\label{eq:matrixM}
	\partial_t\left(\begin{array}{c} \delta\rho_0 \\ \delta\rho_{1}\\\delta\rho_{-1} \end{array}\right) &= {\bf M}\cdot\left(\begin{array}{c} \delta\rho_0 \\ \delta\rho_{1}\\\delta\rho_{-1} \end{array}\right),
\end{align}
where the matrix $\bm M$ is given in  Appendix~\ref{app:LSA}. As shown in Fig.~\ref{Fig:EV}, the largest eigenvalue $\lambda_{\rm max}$ of $\bm M$ may become positive, thus signalling an instability. For $\bm q$ along $\bm e_x$, the eigenmode associated to $\lambda_{\rm max}$ is a linear combination of $(1,0,0)^T$ and $-i(0,1,1)^T$. Physics-wise, this expresses that the particles converge to regions of space where they begin to build up clusters. In the limit where $L_z$ is much larger than the colloidal layer of thickness $2a$, we obtain the instability threshold
\begin{align}
	\left(\mu c_\infty\right)_{\rm th}&\simeq-\frac{2\pi D_t}{1-\frac{\bar\rho_0}{\bar\rho_m}}\left(1+\frac{ L_z}{L_0}\right)\left(1+\frac{L_0}{L_z}\right),\label{eq:muc0crit}
\end{align}
where the characteristic length $L_{0}=D/(h\bar\rho_0)$ is of the order of $L_z$. It depends only on the product $\mu c_\infty$, showing that an increase of the fuel concentration has the same effect as an increase of the phoretic mobility. This is a direct consequence of the fact that self-propulsion is generated by the local chemical field rather than prescribed through a constant propulsion speed.

The apparent divergence at $L_z \to 0$ is avoided since $L_z$ is always larger than the colloid diameter.
When $L_z$ is comparable to this scale, a correction to the threshold arises from the reduced diffusivity in the colloidal layer (see Eq.~\eqref{eq:correction} in Appendix~\ref{app:LSA}).
Conversely, as $L_z \to \infty$, the threshold diverges. This occurs because, for a finite $D$ with the reservoir infinitely far away, the colloids completely deplete the solute, thereby halting chemotaxis.

Remarkably, the threshold is independent of the rotational mobility $\mu'$, indicating that chemotactic torques do not influence the instability threshold to leading order in $\varepsilon$.

We shall use the results of the linear stability analysis to guide our understanding of the particle based numerical results. 

\subsection{Numerical results}
    The JABP model in Eqs.~\eqref{eqa}-\eqref{eqc} is implemented numerically according to Sec.~\ref{sec:Model_exp} in a three-dimensional geometry designed to reproduce the experimental setup ~\cite{theurkauff_dynamic_2012}. The simulation box is filled with a chemical species at a prescribed reservoir concentration, while the Janus particles are confined to the bottom surface. The chemical concentration field evolves dynamically in response to diffusion and solute consumption. Periodic boundary conditions are imposed in the lateral directions, with a no-flux condition at the bottom and a fixed-concentration reservoir at the top. Details of the numerical implementation are provided in Appendix~\ref{app:Numsim}.

\begin{figure}[t!]\centering
{\includegraphics[width=\columnwidth]{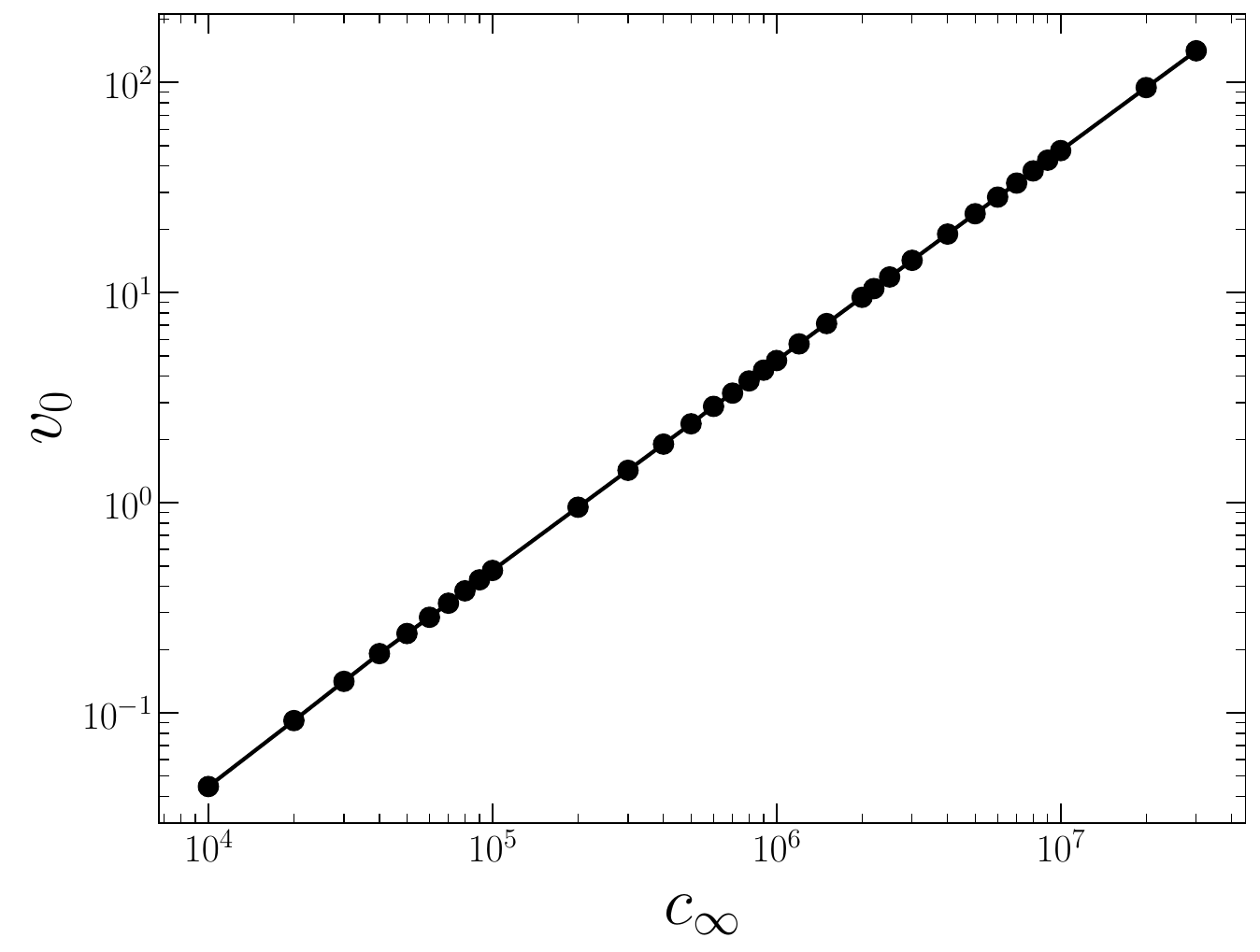}}
\caption{Velocity $v_0$ of a JABP particle as a function of the reservoir concentration $c_\infty$ for $\mu=-10^{-4}$, obtained from a single particle without noise. Simulation parameters are given in Appendix~\ref{app:sim_para}.
}
 \label{Fig:velocity}
\end{figure}

At large $D$, the model predicts  an emergent self-propulsion velocity given by Eq.~\eqref{eq:v0}, which is linear in the chemical concentration. For the simulation of a single noiseless particle at fixed $\mu$, the emergent velocity is found to vary linearly with $c_\infty$, as shown in Fig.~\ref{Fig:velocity}. The theoretical expression is quantitatively reproduced by using $\varepsilon=0.81$, compared with $\varepsilon=0.5$ in the simulations. We attribute the difference to the coarse discretization of the chemical field and to the difference between the imposed reservoir concentration $c_\infty$ and the effective concentration sampled at the particle.

\begin{figure}[t!]\centering
{\includegraphics[width=\columnwidth]{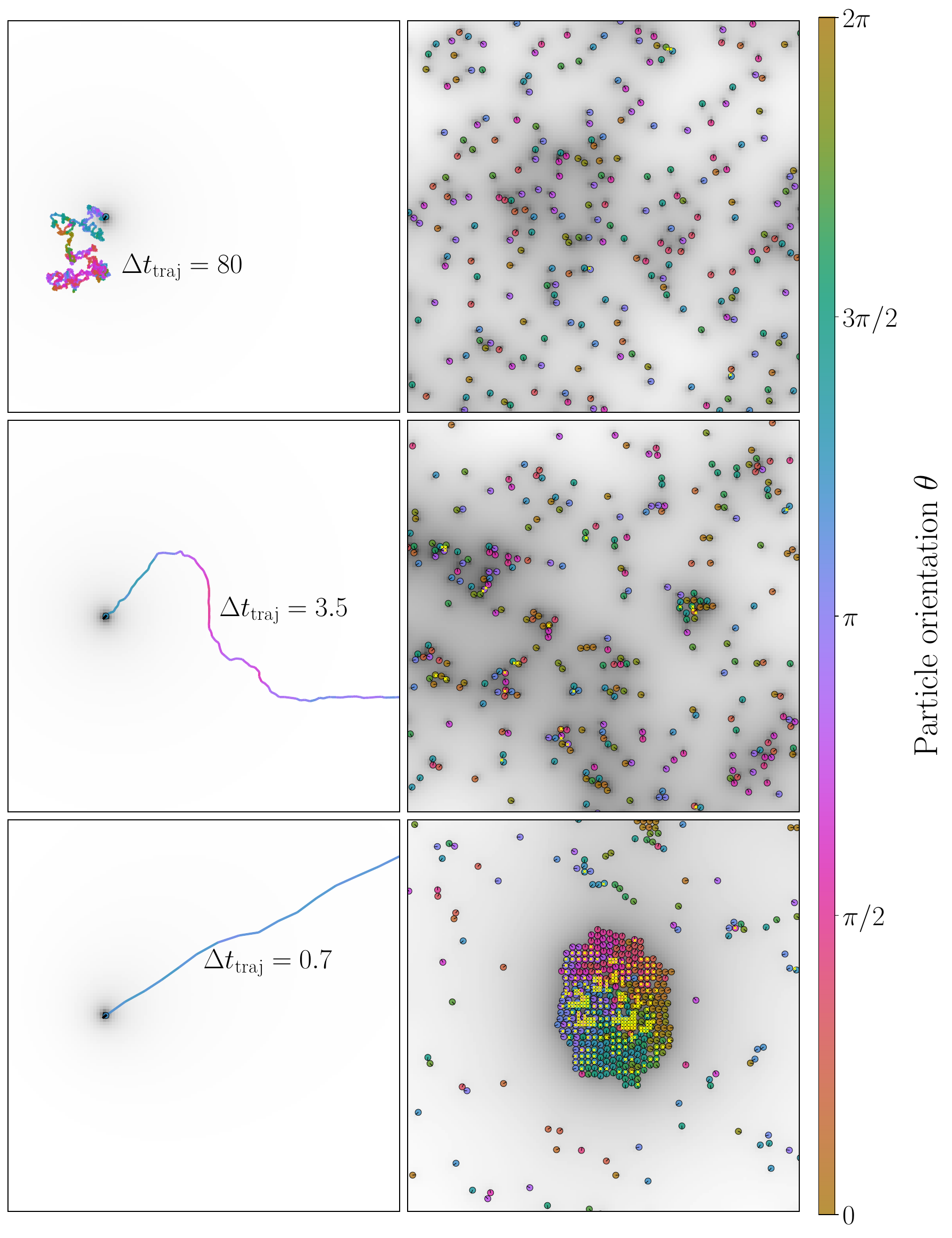}}
\caption{Trajectory of a single particle over a time interval $\Delta t_{\rm traj}$ (left) and particle configurations at a surface packing fraction of $5\%$ (right) for the gas (top), dynamic-clustering (middle), and collapse (bottom) phases. The snapshots show a quarter of the simulation box in the $z=0$ plane. Particle colors indicate their orientations, while yellow crosses mark blocked lattice sites where chemical diffusion is suppressed. The gray levels show the solute concentration. Here, $\mu=-10^{-4}$, $\mu'=10^{-5}$, and $c_\infty=10^4$, $10^7$, and $5\times10^7$ for the gas, dynamic-clustering, and collapse phases, respectively. Other simulation parameters are given in Appendix~\ref{app:sim_para}.}
\label{Fig:Phases}
\end{figure}

\begin{figure}[t!]\centering
{\includegraphics[width=\columnwidth]{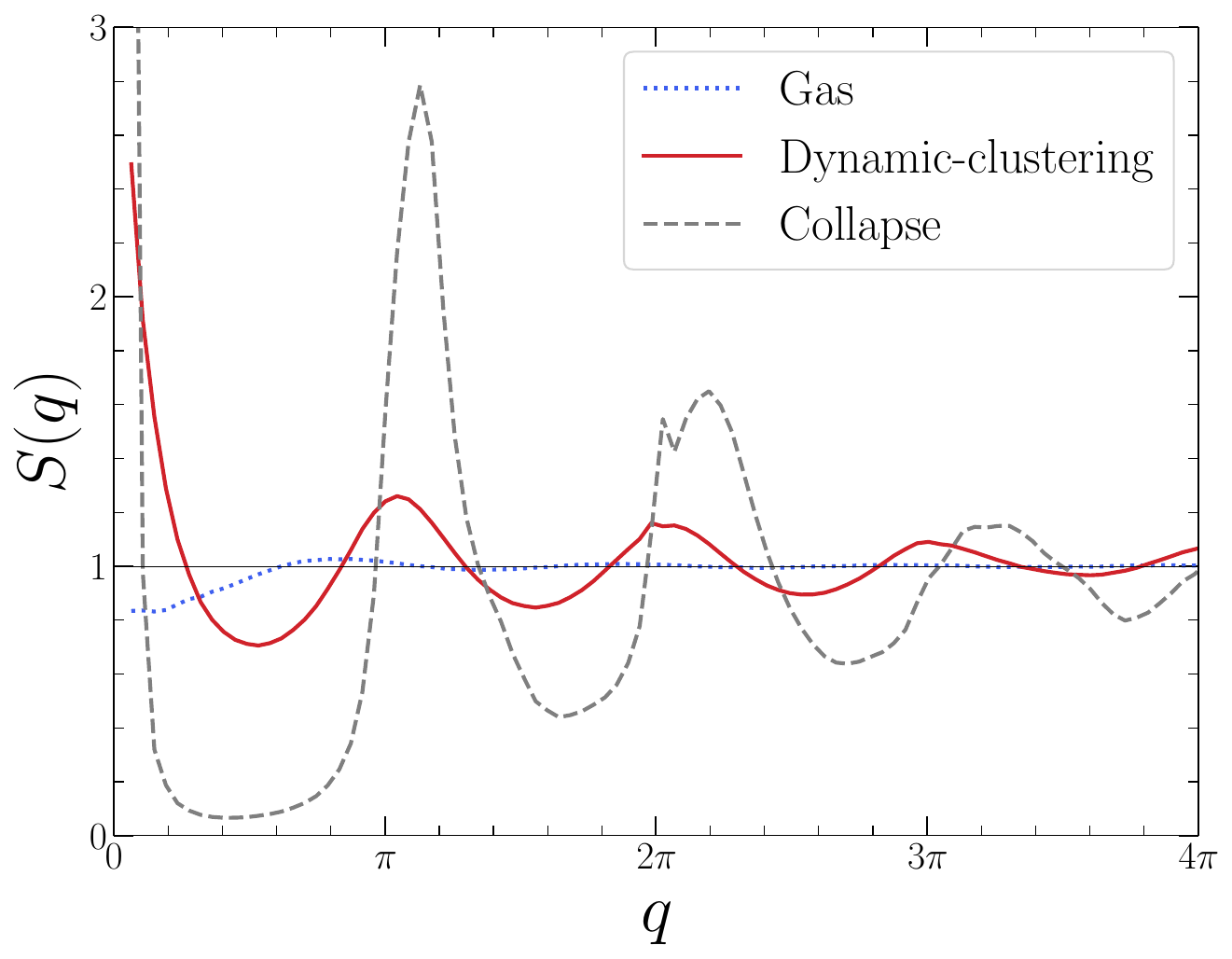}}
\caption{Structure factor $S(q)$ for the gas (blue dotted line), dynamic-clustering (red solid line), and collapse (gray dashed line) phases. The simulation parameters are the same as in Fig.~\ref{Fig:Phases}.}
\label{Fig:SF}
\end{figure}

Turning to the collective dynamics of the system, the experiments reported the formation of dynamic clusters of Janus particles. We consider a dilute colloidal system confined to the bottom surface, with a surface packing fraction of $5\%$. While keeping the particle parameters $\mu$ and $\mu'$ fixed, we vary the chemical concentration $c_\infty$, which increases the self-propulsion velocity and hence the activity. Three regimes are observed: a homogeneous gas at low $c_\infty$, dynamic clustering at intermediate concentrations, and collapse into a dense state at large $c_\infty$. Representative trajectories and configurations for these three regimes are shown in Fig.~\ref{Fig:Phases}. Furthermore, the structure factor in each of these phases shown in Fig.~\ref{Fig:SF} captures our visual observations (see \cite{de2018static} for a recent discussion in active matter), and qualitatively matches those measured in the three phases~\cite{theurkauff_dynamic_2012}.

For the parameters corresponding to the experimental estimates, the system exhibits the dynamic-clustering regime. There, clusters continuously form and dissolve, with particles both joining and leaving existing clusters. The clusters themselves remain mobile, while particles within them undergo persistent motion with a reduced instantaneous velocity (by an indirect solute-mediated form of quorum sensing), reproducing the characteristic dynamical behavior observed experimentally (see the first movie in \cite{SM}). 

The same sequence of phases upon increasing $c_\infty$ is observed for the other values of the rotational mobility that we considered, $\mu'=0$ and $\mu'=\mu/10$. The phase behavior is also robust to alternative choices of the density-dependent diffusion rule and of the steric interaction model.

\begin{figure}[t!]\centering
{\includegraphics[width=\columnwidth]{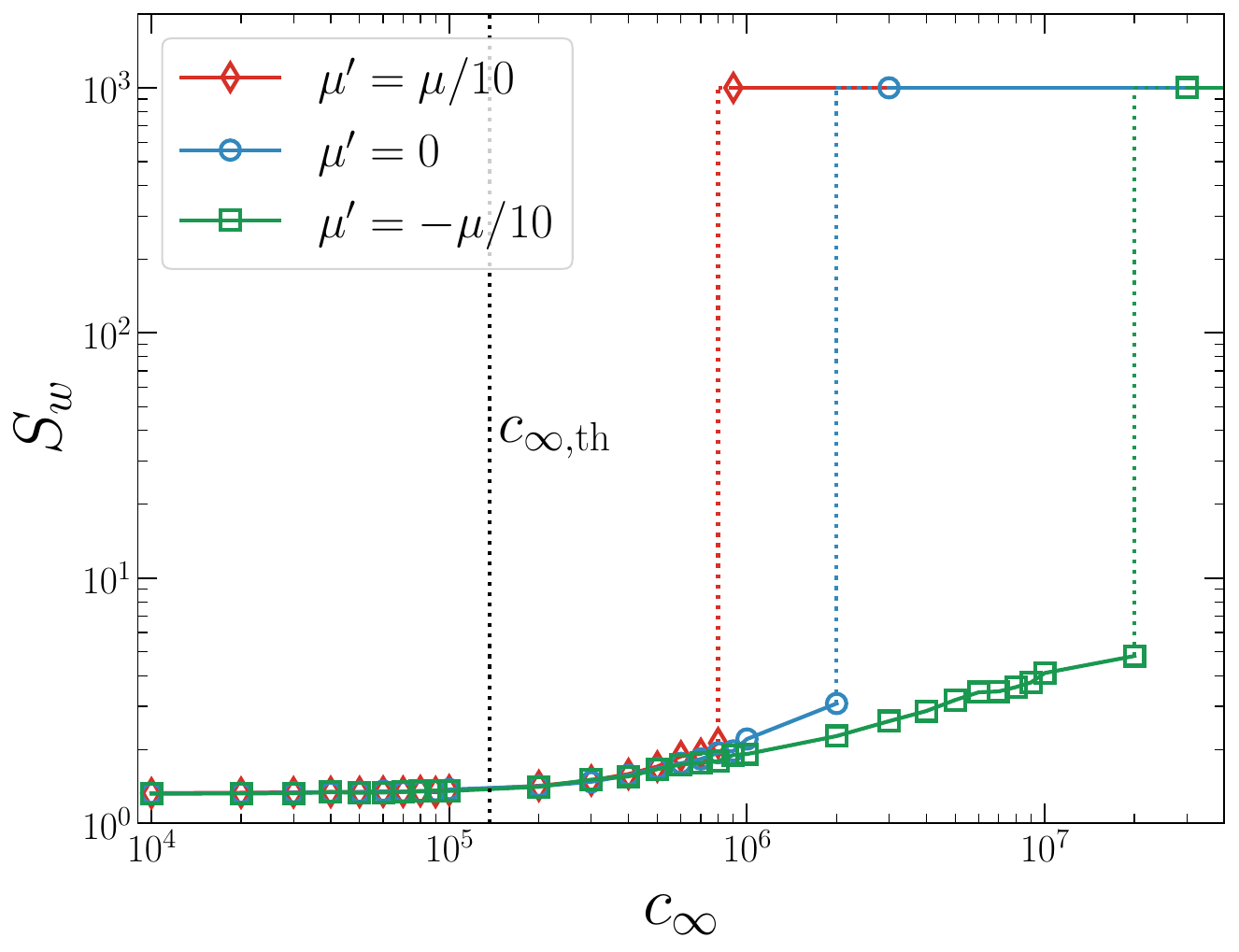}}
\caption{Weighted mean cluster size $S_w$ as a function of the chemical concentration $c_\infty$  for $\mu=-10^{-4}$ and $\mu'=-\mu/10$ (green squares), $\mu'=0$ (blue circles), and $\mu'=\mu/10$ (red diamonds). The theoretical instability threshold $c_{\infty,\mathrm{th}}$ from Eq.~\eqref{eq:muc0crit} is indicated by the vertical gray line. The other vertical dotted lines denote the dynamic-cluster-to-collapse transitions. Other simulation parameters are given in Appendix~\ref{app:sim_para}. }
 \label{Fig:Sw}
\end{figure}

\begin{figure}[t!]\centering
{\includegraphics[width=\columnwidth]{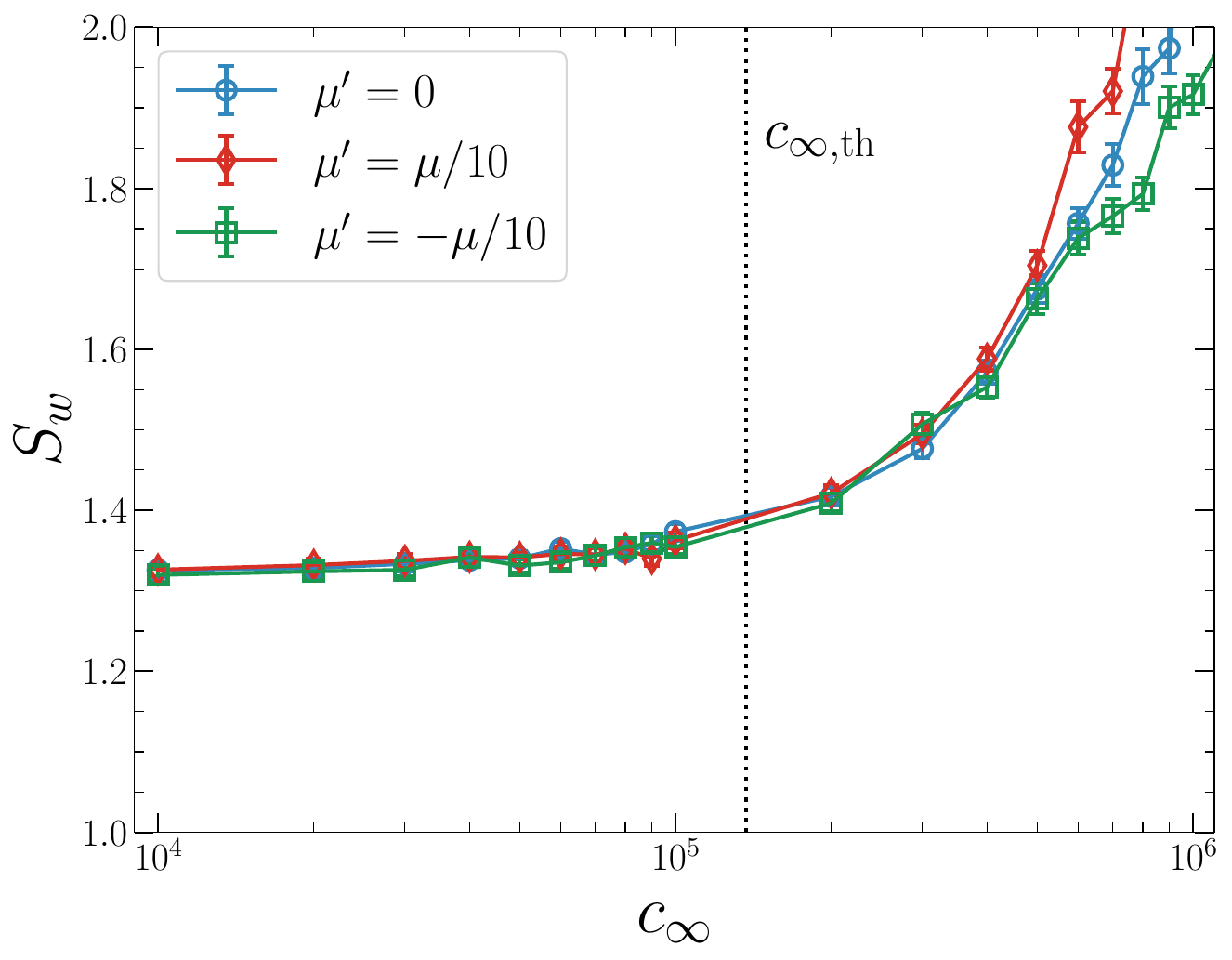}}
\caption{Weighted mean cluster size $S_w$ as a function of the chemical concentration $c_\infty$ near the instability threshold. Symbols and colors are as in Fig.~\ref{Fig:Sw}. The theoretical threshold $c_{\infty,\mathrm{th}}$ from Eq.~\eqref{eq:muc0crit} is indicated by the vertical gray line. Error bars denote the standard error of the estimated averages.} \label{Fig:Sw_zoom}
\end{figure}

Having established that the model reproduces the experimentally observed behavior, we now turn to the prediction of the linear stability analysis. The latter yields an instability threshold given by Eq.~\eqref{eq:muc0crit}. To characterize the transition, we compute the weighted mean cluster size $S_w=\langle{\sum_\alpha N_\alpha^2}\rangle/N,$ where $N_\alpha$ is the number of particles in cluster $\alpha$. Note that $S_w$  corresponds to the average cluster size to which a random particle belongs~\cite{stauffer1994percolation}. Two particles are considered to belong to the same cluster when their center-to-center distance is smaller than three particle radii. The resulting $S_w$ as a function of $c_\infty$ is shown in Figs.~\ref{Fig:Sw} and~\ref{Fig:Sw_zoom}. In the gas phase, $S_w$ remains approximately constant at $S_w\simeq1.3$. Upon crossing the theoretical threshold $c_{\infty,{\rm th}}$, $S_w$ starts to increase, signaling the onset of clustering. The gas-to-dynamic-clustering transition occurs at the same concentration for the three values of $\mu'$. At larger $c_\infty$, however, the onset of collapse depends on $\mu'$, with $S_w$ reaching $N$ at different concentrations. The collapse state, in which all particles belong to a single cluster, is recovered in all cases, as expected. Indeed, at large $c_\infty$, particles creates strong gradients attracting neighboring particles irrespective of their self-propulsion, and thus we have an effective Keller-Segel phenomenology.

\begin{figure}[t!]\centering
{\includegraphics[width=\columnwidth]{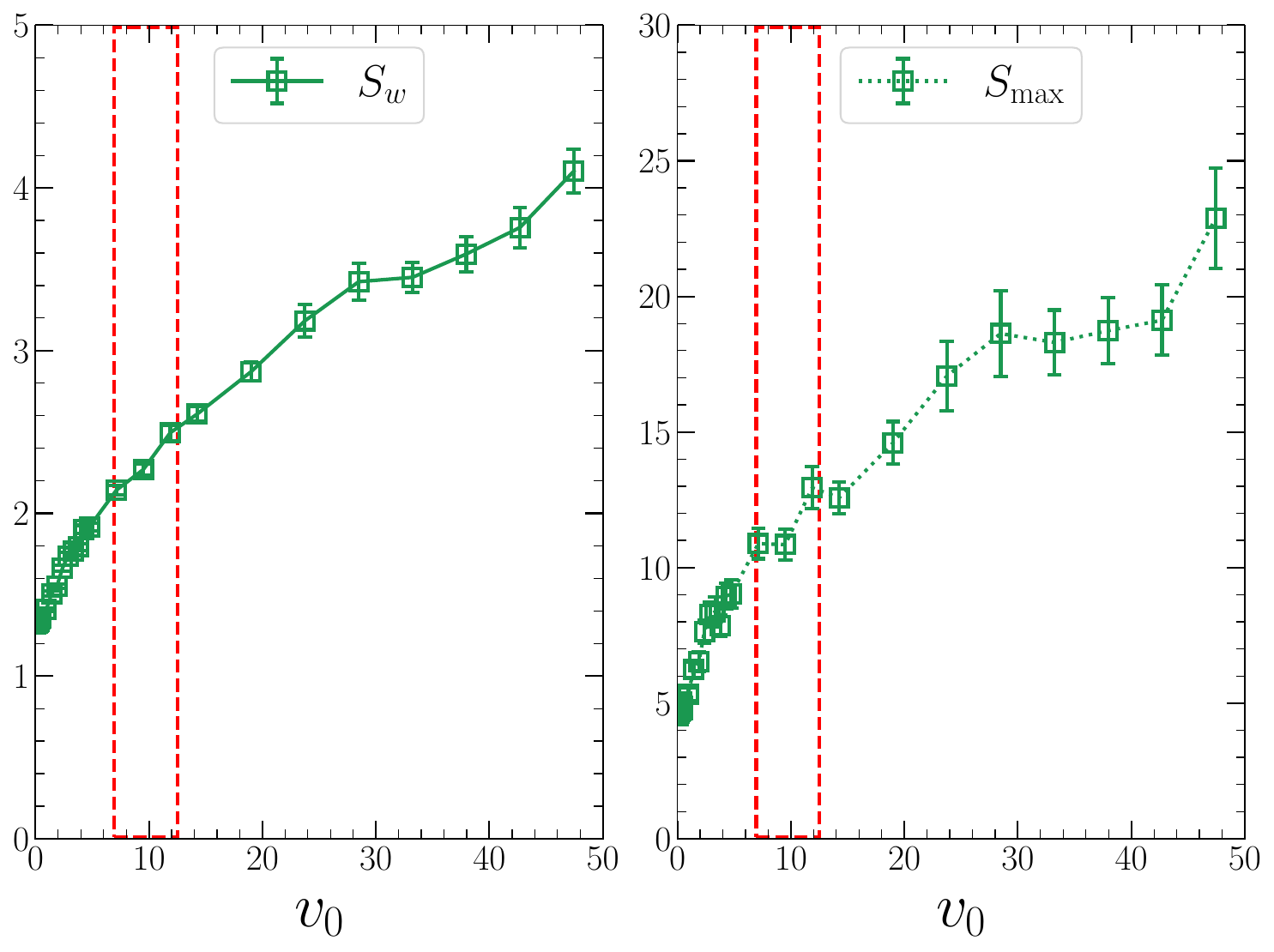}}
\caption{{Weighted mean cluster size $S_w$ (left) and averaged maximal cluster size $S_{\rm max}$ (right) as a function of the self-propulsion velocity $v_0$ for $\mu=-10^{-4}$ and $\mu'=10^{-5}$. Error bars denote the standard error of the estimated averages. The red  dashed rectangle indicates the range of velocities corresponding to the experimental regime. Other simulation parameters are given in Appendix~\ref{app:sim_para}.} } \label{Fig:Sw_v0}
\end{figure}

The experiments reported an approximately linear increase of the mean cluster size with the self-propulsion velocity. As shown on Fig.~\ref{Fig:Sw_v0}, the simulations do not yield a strictly linear dependence on $v_0$ of $S_w$ {and $S_{\rm max}$, the latter being the averaged maximal cluster size}.  However, over the velocity range corresponding to the experiments (red rectangle in Fig.~\ref{Fig:Sw_v0}), the numerical behavior is consistent with the reported trend. This comparison should be interpreted with caution, as the precise value of $\mu'$ is not known experimentally.

\begin{figure}[t!]\centering
{\includegraphics[width=\columnwidth]{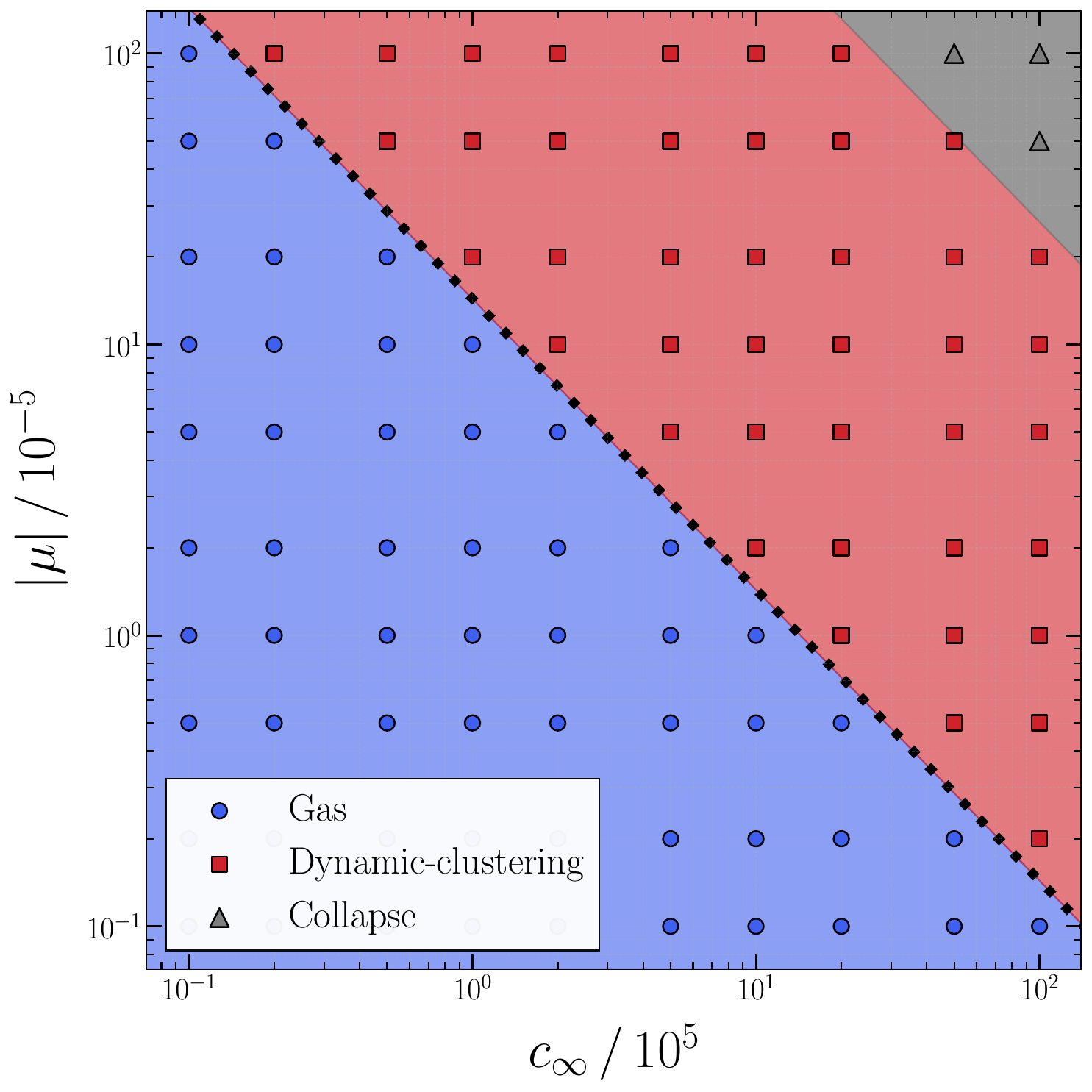}}
\caption{{Phase diagram $|\mu|$ versus $c_\infty$, for $\mu<0$ and $\mu'=-\mu/10$. Symbols represent numerical simulations corresponding to the gas (blue disks in the blue region), dynamic-clustering (red squares in the red region), and collapse (black triangles in the gray region) phases. The black dotted line shows the theoretical threshold $c_{\infty,{\rm th}}$ given in Eq.~\eqref{eq:muc0crit} and found with linear stability analysis.}}
\label{Fig:PD_mu_c0}
\end{figure}

We conclude this section by presenting the phase diagram of the JABP model. We work within the  parameter space spanned by $\mu$, $\mu'$, and $c_\infty$. In Fig.~\ref{Fig:PD_mu_c0}, we show the phase diagram in the $(\mu,c_\infty)$ plane for negative values of $\mu$. The gas to dynamic-clustering transition is in agreement with the theoretical prediction given by Eq.~\eqref{eq:muc0crit}. The slope of the transition line is thus consistent with $\mu\sim c_\infty^{-1}$. The dynamic-clustering to collapse transition exhibits a similar dependence, with $\mu\sim c_\infty^{-1}$ along the transition line.

The location of the gas to dynamic-clustering transition does not depend on the ratio $\mu'/\mu$, in agreement with the theoretical prediction of Eq.~\eqref{eq:muc0crit}. This behavior is consistent with the results of Figs.~\ref{Fig:Sw} and \ref{Fig:Sw_zoom}. The dynamic-clustering to collapse transition, on the other hand, depends on $\mu'/\mu$. It remains parallel to the gas to dynamic-clustering transition, indicating the same $\mu\sim c_\infty^{-1}$ dependence. The width of the dynamic-clustering region therefore varies with $\mu'$, with antialignment favoring a broader dynamic-clustering regime. This behavior can be understood from the fact that dynamic clustering requires the particles to reorient sufficiently rapidly to escape the clusters; when reorientation is too slow, the system undergoes collapse, as already reported in the literature.

We finally consider the phase diagram in the $(\mu,v_0)$ plane. Using Eq.~\eqref{eq:v0}, $c_\infty$ is expressed in terms of $v_0$ and $\mu$. The dynamic-clustering phase extends over a broad range of self-propulsion velocities, spanning nearly two orders of magnitude. The transition lines are vertical, consistent with the scaling $v_0\sim\mu c_\infty$. Combining Eq.~\eqref{eq:v0} with the theoretical prediction for $(\mu c_\infty)_{\rm th}$, Eq.~\eqref{eq:muc0crit}, gives
\begin{align}
v_{0,\mathrm{th}}\simeq  \frac{hD_t}{2D\varepsilon^2}\frac{1}{1-\frac{\bar\rho_0}{\bar\rho_m}}\left(1+\frac{L_z}{L_{0}}\right)\left(1+\frac{L_0}{L_z}\right),\label{eq:v0_th}
\end{align}
which is independent of $\mu$ or $c_\infty$. Regarding the dynamic-clustering to collapse transition, increasing the ratio $\mu'/\mu$ shifts the transition line to the left.

Overall, the gas to dynamic-clustering transition is controlled by the product $\mu c_\infty$, or equivalently by $v_0$, in agreement with our theoretical prediction and our numerical results. Numerical results further show that the dynamic-clustering to collapse transition is controlled by the ratio $\mu'/\mu$.

\begin{figure}[t!]\centering
{\includegraphics[width=\columnwidth]{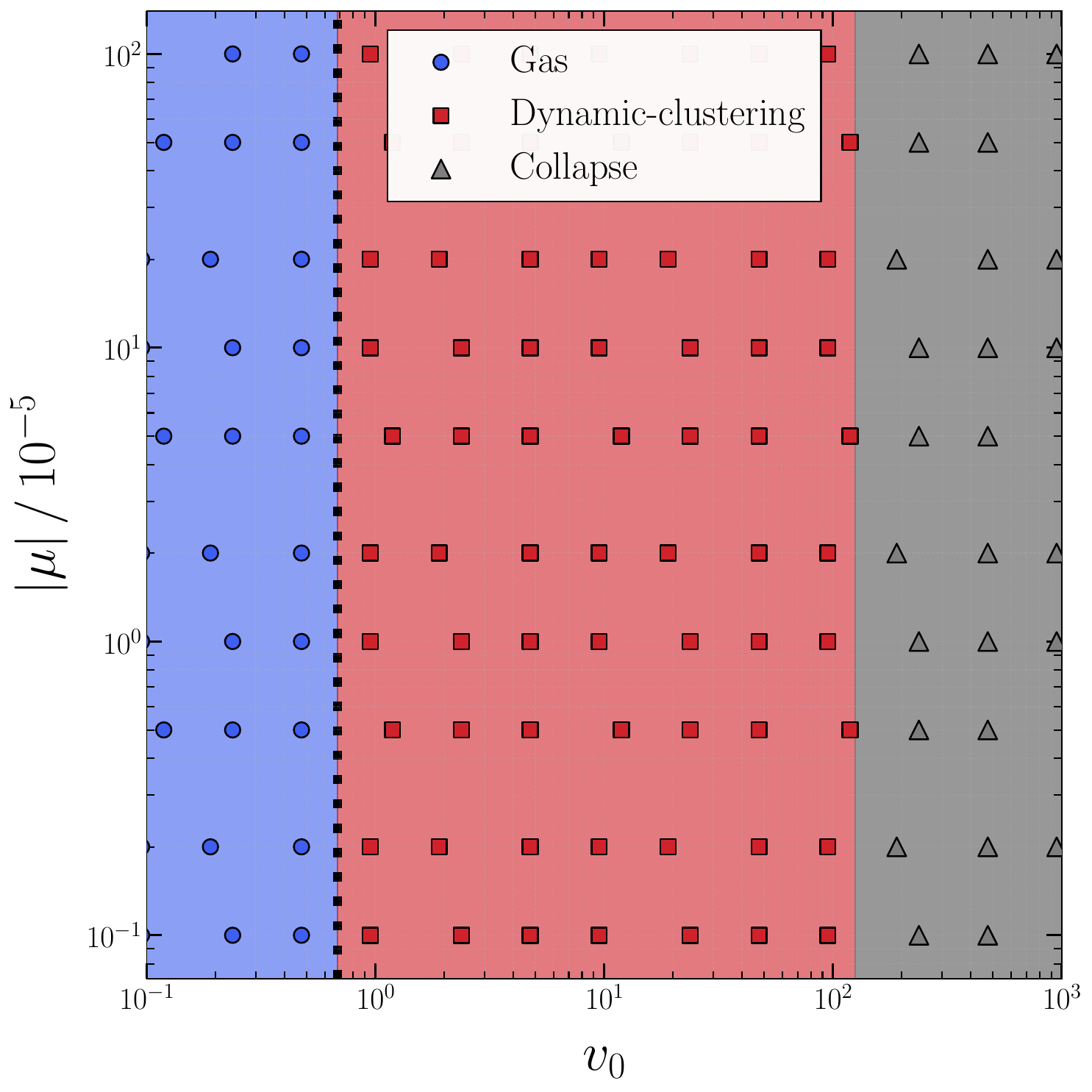}}
\caption{{Phase diagram $|\mu|$ versus $v_0$, for $\mu<0$ and $\mu'=-\mu/10$. Symbols represent numerical simulations corresponding to the gas (blue disks in the blue region), dynamic-clustering (red squares in the red region), and collapse (black triangles in the gray region) phases. The black dotted line shows the theoretical threshold Eq.~\eqref{eq:v0_th}.}}
\label{Fig:PD_mu_v0}
\end{figure}

There are two reasons why a direct comparison between our phase diagram in Fig.~\ref{Fig:PD_mu_v0} and Fig.~4 of \cite{pohl_dynamic_2014} is difficult. While in our setting $v_0$ directly depends on the solvent, the choice made in \cite{pohl_dynamic_2014} is to resort to independent parameters. There phase diagram in the $(v_0,\mu)$ plane at fixed $\mu'$, as they acknowledge, should be studied along a line starting from the origin, since these parameters are all proportional to $c_\infty$. In fact, by doing so, they keep $\mu'$ a constant, whereas it should also be proportional to $c_\infty$. Hence our difficulties for a direct comparison, since our approach integrates these mutual dependencies. Strikingly, this notably simplifies the shape of the phase diagram. Unfortunately, we have neither been been able to locate two dynamical cluster phases, nor even to confirm their existence.

\section{Conclusion}
\label{sec:conclustion}

The model we have proposed for the experimental setup is perhaps more realistic than previously proposed ones. It takes into account the specific three-dimensional geometry, it does not postulate an endogenous self-propulsion but features the latter as an emergent diffusiophoretic process as in real Janus colloids. In addition, the solute concentration field is treated as a dynamical variable while taking into account the hindered diffusion within dense regions, notably inside dense clusters. The main features of the phase diagram of \cite{pohl_dynamic_2014} are robust to our refinements. However, our analysis sheds a new light on the extent of the dynamic cluster phase, which appears to be more than just a pre-transitional effect. The dynamic clustering fraction of the phase diagram is controlled by $c_\infty$, which, within our approach, we can vary as in the experiment and $c_\infty$ also controls the self-propulsion velocity. We see that dynamic cluster exists over one or two orders of magnitude of $c_\infty$, in a fashion robust to changes in the $\mu$ parameter. 

The interpretation of dynamic clustering in terms of an arrested Keller-Segel collapse (by the orientational phoretic mobility favoring the escape of particles out of dense clusters), as already suggested in \cite{pohl_dynamic_2014} is indeed very appealing. We do find that the Keller-Segel equations describe our system in a specific limit. However, what we also see, is that motility-induced phase separation is at work. Increasing $c_\infty$ increases the Péclet number and favors the coarsening of a dense phase. In practice this is achieved both by the in-cluster slow down due to excluded-volume collisions and by the velocity reduction caused by the shortening of solute {(see the second movie in \cite{SM})}. The latter mechanism is equally limited by the orientational phoretic coupling and we find it hard to sort out the dominant one among motility-induced phase separation and Keller-Segel collapse. It would certainly be interesting to seek a protocol ({\it in silico} or {\it in vivo}) allowing us to answer this open question.

\acknowledgements We thank Mike  Cates, Ananyo  Maitra, Kranthi Mandadapu,  Giulia Pisegna, Sriram Ramaswamy and Julien Tailleur for very useful discussions. {We also thank Eliot Selmer for his assistance with preliminary numerical analysis}. JBF and FvW acknowledge the hospitality of the Kavli Institute for Theoretical Physics (KITP), University of California, Santa Barbara, during which part of this work was elaborated.


\appendix

\section{Dean-Kawasaki equation}
\label{app:DK}

Using the It\=o convention, the variations of $\bm r_n$ and $\theta_n$ between  $t$ and $t+dt$, for the Janus-ABP model in $d=2$ dimensions, are given by
\begin{align}
d\bm r_n&=\mu\bm\nabla c(\bm r_n)dt+\sqrt{2 D_t}\,d\bm W_n,
\nonumber\\
d\theta_n&=\mu'\bm u^\perp_n\cdot\bm\nabla c(\bm r_n)dt+\sqrt{2D_r}\,d\hat W_n,
\end{align}
where $d\bm W_n=\int_t^{t+dt}\bm \xi_n(t')dt'$ and $d\hat W_n=\int_t^{t+dt}\eta_n(t')dt'$ are uncorrelated Wigner processes, of null average and variance equal to $dt$.

Let us first calculate the first two moments of the variation $d\rho=\rho(\bm r,\theta,t+dt)-\rho(\bm r,\theta,t)$. Because $\rho=\sum_n\delta(\bm r-\bm r_n)\delta(\theta-\theta_n)$ is a function $\sum_n\psi(\bm r_n,\theta_n)$ of stochastic variables, we compute $d\rho$ using It\=o's lemma (second order Taylor expansion):
\begin{align}
    d\rho=\sum_n&\bigg[
    \frac{\partial\psi}{\partial r_{ni}}dr_{ni}
    +\frac12\frac{\partial^2\psi}{\partial r_{ni}\partial r_{nj}}dr_{ni}dr_{nj}
    \nonumber\\
&+\frac{\partial\psi}{\partial\theta_n}d\theta_n
+\frac12\frac{\partial^2\psi}{d\theta_n^2}d\theta_n^2
+\frac{\partial^2\psi}{dr_{ni}d\theta_n}dr_{ni}d\theta_n
    \bigg],
\end{align}
where all variations are calculated at time $t$ according to the It\=o convention. Taking the statistical average, we get
\begin{align}
    &\lim_{dt\to0}\frac{\langle d\rho\rangle}{dt}
    = - \bm \nabla\cdot
    \left(
    \rho\mu\bm\nabla c
    -D_t\bm\nabla\rho
    \right)
    -\partial_\theta\left(
    \rho\mu'\bm u^\perp
    \cdot\bm\nabla c
    \right),
\\
   &\lim_{dt\to0}\frac{\langle d\rho\,d\rho'\rangle}{dt}
    =
    2(D_t\partial_i\partial'_i
    +D_r\partial_\theta\partial_\theta')
    \nonumber\\&
    \hspace{3cm}
    \left[
    \rho(\bm r)\delta(\bm r-\bm r')\delta(\theta-\theta')
    \right],
\end{align}
where $d\rho'$ is calculated at point $(\bm r', \theta')$ and $\partial'_i=\partial/\partial r'_i$.

Calculating these quantities from the Dean-Kawasaki equation
\begin{align}\label{eq:DKAppendix}
    \partial_t\rho &= - \bm \nabla\cdot
    \left(
    \rho\mu\bm\nabla c
    -D_t\bm\nabla\rho
    -\sqrt{2D_t\rho}\,\bm \xi
    \right)
    \nonumber\\
    &-\partial_\theta\left(
    \rho\mu'\bm u^\perp
    \cdot\bm\nabla c
    -D_r\partial_\theta\rho
    -\sqrt{2D_r\rho}\,\eta
    \right),
\end{align}
with delta-correlated Gaussian noises $\bm\xi$ and $\eta$ and the It\=o convention, gives the same results. This shows the correctness of Eq.~\eqref{eq:DKAppendix}, since higher moments vanish at order $dt$.

\section{Linear-stability analysis}
\label{app:LSA}
In this Appendix we derive the instability threshold reported in the main text. 

\paragraph{Background state.}

We consider a stationary isotropic   homogeneous state with particle density $\bar\rho_0 = N/S$, where $S$ is the container area. The associated concentration profile $c_0(z)$ varies with $z$ due to the reservoir at $L_z$, the no-flux boundary condition at the bottom, solute depletion by the colloids, and the reduced diffusivity within the colloid layer. Regarding the latter, a finite thickness of the colloid layer is essential for the spatial dependence of the diffusivity to take effect.

From Eq.~\eqref{eq:cepsilon}, the background concentration profile $c_0(z)$ obeys the equation
\begin{align}
D\partial_z\left[\left(1-{g(z)\bar\rho_0}/{\bar{\rho}_m}\right)\partial_z  c_0\right]-h c_0(z)\delta(z)\bar{\rho}_0
=0,\label{eq:eomc_0}
\end{align}
with $g(z)=1$ for $z\le2a$ and $g(z)=0$ for $z>2a$, where $a$ is the colloid radius.
We solve it separately in the regions $z\leq2a$ and $z>2a$, imposing continuity of both the concentration field and the diffusive flux at the interface,
\begin{align}
D\left.\partial_z c\right|_{z\to(2a)^+}
=
D\left(1-\bar\rho_0/\bar\rho_m\right)
\left.\partial_c\right|_{z\to(2a)^-},
\end{align}
yielding the concentration profile $ c_0(0) $ in $z=0$ given in Eq.~\eqref{eq:c00}.  

\paragraph{Linear stability analysis.}

In the fast diffusion limit, the Fourier modes of the density and concentration perturbations $\delta\rho(\bm r,\theta,t)$ and $\delta c(\bm r,\theta,t)$, defined in the main text, are solutions of Eq.~\eqref{eq:cepsilon} with $\partial_t\delta c=0$. Hence, at linear order, they obey of the coupled set of equations
\begin{align}
    \partial_t\delta\rho_0 &= \frac{\bar\rho_0}{2\pi}\left(1-{\bar{\rho}_0}/{\bar{\rho}_m}\right)\mu q^2\delta c_s-D_tq^2\delta\rho_0,\\
	\partial_t\delta\rho_{\pm 1} &= -(D_t q^2  +D_r)\delta\rho_{\pm 1}+\frac{\mu'\bar\rho_0}{4\pi}q_\pm\delta c_s,\\
	\partial_t\delta\rho_{m} &= -\frac{1}{2}(D_t q^2  +D_r m^2)q^2 \delta\rho_{m},\\
	0 & =-q^2D\left(1-{g(z)\bar\rho_0}/{\bar{\rho}_m}\right) \delta c\notag\\
  & +  \partial_z\left[ D\left(1-{g(z)\bar\rho_0}/{\bar{\rho}_m}\right)\partial_z \delta c \right] -h\bar \rho_0 \delta(z) \delta c \notag\\
    & +h  c_0\delta(z)\left(-\delta \rho_0+\frac{\epsilon}{2}\delta\rho_{1}q_-+\frac{\epsilon}{2}\delta\rho_{-1}q_+\right),\label{eq:eomdc}
\end{align}
where $q_\pm = iq_x\pm q_y$, $m\neq0,\pm1$, and $\delta c_s(\bm q)=\delta c(\bm q,z=0)$. The higher angular modes $|m|\ge2$ are decoupled from the density and polarization modes and therefore play no role in the linear instability.

Solving Eq.~\eqref{eq:eomdc} gives
\begin{align}
	\delta c_s &=\frac{h c_0(0)\left(-\delta \rho_0+\frac{\epsilon}{2}\delta\rho_{1}q_-+\frac{\epsilon}{2}\delta\rho_{-1}q_+\right)}{ h \bar{\rho}_0+D q \coth(L_z q) },\label{eq:dcs}
\end{align}
which we substitute in the linearized equations for $\delta\rho_0, \delta\rho_{1}$, and $\delta\rho_{-1}$. The evolution matrix $\bm M$ defined in Eq.~\eqref{eq:matrixM} is such that
\begin{align}
	&-\bf M=\notag\\
	&\left(
	\begin{array}{ c c c} D_t q^2+\frac{\mu_{\rm eff}}{J}q^2 & -\frac{\mu_{\rm eff}\epsilon}{2J}q^2q_- & -\frac{\mu_{\rm eff}\epsilon}{2J}q^2q_+\\
	\frac{\mu'}{2J}q_+ & D_r(q)+\frac{\mu'\epsilon}{4J}q^2& -\frac{\mu'\epsilon}{4J}q_+^2\\
	\frac{\mu'}{2J}q_- &  -\frac{\mu'\epsilon}{4J}q_-^2 & D_r(q)+\frac{\mu'\epsilon}{4J}q^2
	\end{array}
	\right)
\end{align}
where $D_r(q)= D_r+D_tq^2$
and
\begin{align}
	\mu_{\rm eff}& = \mu\left(1-\frac{\bar\rho_0}{\bar\rho_m}\right),\\
	J(q)&=\frac{2\pi}{c_\infty}\left[1+h\bar\rho_0\left(\frac{2a}{D\left(1-\bar\rho/\bar\rho_m\right)}+\frac{L_z}{D}\right)\right]\times\notag\\
	&\left[1+\frac{Dq}{h\bar\rho_0}\coth(L_zq)\right].
\end{align}

The eigenvalues of ${\bf M}$ are
\begin{align}
	{\boldsymbol \lambda} &=\left\{-D_r-D_tq^2,\right.\notag\\
	&-\frac{D_r}{2}-{D_t}q^2-\frac{1}{4}q^2\frac{2\mu_{\rm eff}+\epsilon\mu'}{J}\notag\\
	&\left.\pm\frac{1}{4}\sqrt{4D_r^2+4D_rq^2\frac{-2\mu_{\rm eff}+\epsilon\mu'}{J}+q^4\frac{(2\mu_{\rm eff}+\epsilon\mu')^2}{J^2}}\right\}.
\end{align}
Only the eigenvalue with plus sign in front of the square root can become positive, signaling the onset of the instability.

The instability threshold is obtained by imposing that this eigenvalue vanishes in the long-wavelength limit, giving
\begin{align}\label{eq:correction}
	&\left(\mu c_\infty\right)_{\rm th}=\notag\\
	&-2\pi D_t\frac{1+\frac{h\bar\rho_0 L_z}{D}}{1-\frac{\bar\rho_0}{\bar\rho_m}}\left[1+\frac{D}{h\bar\rho_0L_z}+\frac{2a}{L_z\left(1-\frac{\bar\rho_0}{\bar\rho_m}\right)}\right],
\end{align}
and Eq.~\eqref{eq:muc0crit} is recovered by neglecting the last term on the right-hand side, since $a\ll L_z$.

\section{Numerical simulations}
\label{app:Numsim}
In this Appendix, we describe the numerical implementation of the Janus model used in Sec.~\ref{sec:Model_exp}.

\subsection{Simulation setup.}

We consider a three-dimensional cubic box with $L$ lattice sites along the $x$ and $y$ directions and $L_z$ lattice sites along the $z$ direction. Lattice sites are labeled by $(i,j,k)$, where $0\le i,j<L$ and $0\le k<L_z$. The lattice spacing is chosen equal to the particle radius $a$ and, throughout the simulations, we set $\epsilon=a$. Ideally, we should work with a much finer discretization grid, which would notably slow the simulations for no-to-little physical gain, hence our minimalist choice of the lattice spacing.

The chemical field occupies the entire three-dimensional lattice and is defined on the lattice sites. A no-flux boundary condition is imposed at the bottom boundary, between the physical layer with index $k=0$ and the fictitious layer $k=-1$. At the top boundary, a reservoir of fixed concentration $c_\infty$ is imposed through the fictitious layer $k=L_z$, such that $c_{i,j,L_z}=c_\infty$ for all $i$ and $j$. The chemical concentration and its gradient at arbitrary positions in the plane $z=0$ are obtained by bilinear interpolation from the four nearest lattice sites.

The $N$ particles are confined to the plane $z=0$, corresponding to the lattice layer $k=0$. Their positions $\mathbf r_n\in[0,L[^2$ and in-plane orientations $\theta_n\in[0,2\pi[$ evolve continuously, while periodic boundary conditions are applied along the $x$ and $y$ directions.

Initially, particle positions are drawn uniformly in the $(x,y)$ plane, subject to the excluded-volume constraint, and particle orientations are sampled from a uniform distribution  on $[0,2\pi[$. The chemical field is initialized uniformly with concentration $c_\infty$ over the whole the simulation box.

\subsection{Time-stepping algorithm.}
Each simulation time step consists of three successive stages:
(i) diffusion of the chemical field,
(ii) chemical consumption by the particles,
and (iii) update of particle positions and orientations.

\paragraph{Chemical field diffusion.}
The chemical field diffuses within the three-dimensional volume. Standard diffusion is applied in the bulk ($k>0$), while the reservoir boundary condition is implemented by allowing diffusion from the fictitious layer $k=L_z$, whose concentration is kept fixed at $c_\infty$.

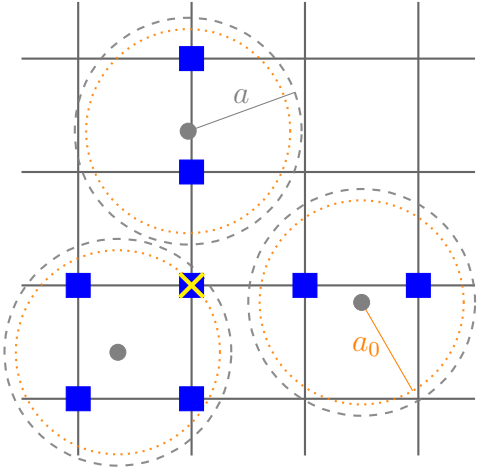
\begin{figure}[t!]
    \centering
\begin{tikzpicture}[scale=1.5]
	\draw[draw=black!60, thick] (0.5, 0) -- (0.5, 4);    
	\draw[draw=black!60, thick] (1.5, 0) -- (1.5, 4);    
	\draw[draw=black!60, thick] (2.5, 0) -- (2.5, 4);    
	\draw[draw=black!60, thick] (3.5, 0) -- (3.5, 4);       	
	\draw[draw=black!60, thick] (0, 0.5) -- (4, 0.5);
	\draw[draw=black!60, thick] (0, 1.5) -- (4, 1.5);
	\draw[draw=black!60, thick] (0, 2.5) -- (4, 2.5);
	\draw[draw=black!60, thick] (0, 3.5) -- (4, 3.5);
	\filldraw[draw=gray, fill=gray] (0.85,0.91) circle (2pt);
	\draw[draw=gray, thick, dashed, opacity=0.9] (0.85,0.91) circle (1.);
	\draw[draw=orange, thick, dotted, opacity=0.9] (0.85,0.91) circle (.9);
	\filldraw[draw=gray, fill=gray] (1.47,2.86) circle (2pt);
	\draw[draw=gray, thick, dashed, opacity=0.9] (1.47,2.86) circle (1.);
	\draw[draw=orange, thick, dotted, opacity=0.9] (1.47,2.86) circle (.9);
	\draw[draw=gray] (1.47,2.86) -- ++(20:1) node[midway, above,  text=gray,font=\large] {$a$};	
	\draw[draw=gray, thick, dashed, opacity=0.9] (3.0,1.35) circle (1.);
	\draw[draw=orange, thick, dotted, opacity=0.9] (3.0,1.35) circle (.9);
	\draw[draw=orange] (3.0,1.35) -- ++(300:0.9) node[midway, left=-1pt,  text=orange,font=\large] {$a_0$};
	\filldraw[draw=gray, fill=gray] (3.0,1.35) circle (2pt);
	\filldraw[draw=blue, fill=blue] (0.5,0.5) ++(-3pt,-3pt) rectangle ++(6pt,6pt);
	\filldraw[draw=blue, fill=blue] (1.5,0.5) ++(-3pt,-3pt) rectangle ++(6pt,6pt);
	\filldraw[draw=blue, fill=blue] (0.5,1.5) ++(-3pt,-3pt) rectangle ++(6pt,6pt);	
	\filldraw[draw=blue, fill=blue] (1.5,1.5) ++(-3pt,-3pt) rectangle ++(6pt,6pt);
	\draw[yellow, ultra thick] (1.5,1.5) ++(-3pt,-3pt) -- ++(6pt,6pt);
	\draw[yellow, ultra thick] (1.5,1.5) ++(-3pt,3pt) -- ++(6pt,-6pt);
	\filldraw[draw=blue, fill=blue] (2.5,1.5) ++(-3pt,-3pt) rectangle ++(6pt,6pt);
	\filldraw[draw=blue, fill=blue] (3.5,1.5) ++(-3pt,-3pt) rectangle ++(6pt,6pt);
	\filldraw[draw=blue, fill=blue] (1.5,2.5) ++(-3pt,-3pt) rectangle ++(6pt,6pt);
	\filldraw[draw=blue, fill=blue] (1.5,3.5) ++(-3pt,-3pt) rectangle ++(6pt,6pt);
\end{tikzpicture}
\caption{Illustration of the density-dependent diffusion rule used in the simulations.
The underlying square lattice represents the grid on which the chemical field is defined,
with lattice spacing equal to the particle radius $a$.
Particles have continuous positions, indicated by gray disks, and their physical radius
is represented by the dashed gray circle of radius $a$.
A lattice site is considered occupied if it lies within the orange dotted circle of radius
$a_0=0.9a$ around a particle center. Occupied lattice sites are shown as blue squares.
An occupied site is classified as blocked (yellow cross) when it has more than two occupied
nearest lattice neighbours. Chemical diffusion to and from blocked sites is then suppressed.}
    \label{fig:dddiff}
\end{figure}

To mimic the reduction of chemical diffusion inside dense clusters of particles, diffusion within the first layer ($k=0$) is made configuration-dependent in the following way.  Each lattice site is classified as either \emph{free} or \emph{blocked} regarding diffusion. If the discretization grid were chosen with a much finer step, things would be simple: sites would be blocked up to a distance $a$ from the particle center. Due to our coarse grid, we have adapted the blocking criterion as follows.  A site is considered occupied if it lies within a distance $a_0$ from the center  of any particle. We choose $a_0<a$ in order to account for the actual diffusion through the interstices around the particles. An occupied site is classified as blocked if more than two of  its nearest lattice neighbours in the $(x,y)$ plane are also occupied. In this sense, the diffusivity depends on the local density: it is simply reduced in sparse clusters and it vanishes in close-packed clusters, in agreement with the continuous model. The resulting rule is illustrated in Fig.~\ref{fig:dddiff}. The diffusion process is then defined according to the following rules:
\begin{itemize}
\item No diffusion is allowed to or from the fictitious layer $k=-1$, implementing the no-flux boundary condition.
\item No diffusion is allowed to or from blocked sites.
\item The concentration at the reservoir layer $k=L_z$ is maintained at the constant value $c_\infty$, providing a constant supply of chemical by diffusion.
\end{itemize}

\paragraph{Chemical consumption.}
For each particle, the consumption point {$\bm r_n+\epsilon\bm u_n$} is determined from its position and orientation according to the JABP model. Consumption always occurs in the layer $k=0$ {since $\bm u_n$ lies within the bottom plane}. The amount of chemical consumed is proportional to the local concentration, obtained by bilinear interpolation from the four nearest lattice sites. The concentration at these sites is then updated accordingly, irrespective of whether they are classified as free or blocked. 

\paragraph{Particle dynamics.}
The chemical concentration and its gradient at the particle positions are obtained by bilinear interpolation from the four nearest lattice sites. Particle positions and orientations are then updated according to the stochastic equations of motion.

Steric interactions are modeled through a soft repulsive potential. Whenever two particles are separated by a center-to-center distance $r<2a$, the particle equation of motion includes the repulsive drift  force
\begin{align}
\bm f_{\rm ex}=K(2a-r)\hat{\bm r},
\end{align}
where $\hat{\bm r}=\bm r/r$ is the unit vector joining the particle centers and $K$ controls the repulsion strength.\\

\subsection{Dimensionless parameters.}
\label{app:sim_para}

The simulations are performed using dimensionless variables. Lengths and times are measured in units $\ell_0=\sqrt{{D_t}/{D_r}}$, and $t_0=1/(2D_r)$, while the chemical concentration is measured in units of $\ell_0^{-3}$. With this choice, the dimensionless diffusion coefficients become
\begin{align}
D_t=D_r=\frac12.
\end{align}
Using the experimental values reported by Theurkauff \textit{et al.}~\cite{theurkauff_dynamic_2012}, these units correspond to $\ell_0\simeq\SI{1.8}{\micro\meter}$, and $t_0\simeq\SI{5}{\second}$.
We choose the values of $\mu$ and $h$ in order to recover the velocity $v_0 \simeq \qty{3}{\micro\metre\per\second}$ measured experimentally for the concentration $c_\infty=\qty{e-4}{\mol\per\liter}$. We choose $K$ in order for particles to exhibit only little penetration over at most one time step. This yields
\begin{align}
D&=1000, &
\mu&=-10^{-4}, &
h&=400,\notag\\
a&=\epsilon=0.5, &
a_0&=0.9a, &
K&=10^{4}.
\end{align}
Unless stated otherwise, we use these values throughout.
The rotational mobility is taken as
\begin{align}
\mu'\in\left\{\frac{\mu}{10},\,0,\,-\frac{\mu}{10}\right\},
\end{align}
while the reservoir concentration $c_\infty$ is used as the control parameter.

The numerical simulations are performed with
\begin{align}
N&=1000, &
L&=250, &
L_z&=100,
\end{align}
corresponding to an initial surface packing fraction of approximately $5\%$.

\bibliography{JanusABP}

\end{document}